\documentclass[]{spie}  

\usepackage{amsmath,amsfonts,amssymb}
\usepackage{graphicx}
\usepackage[colorlinks=true, allcolors=blue]{hyperref}
\usepackage{array}

\title{MOSAIC: the high-multiplex and multi-IFU spectrograph for the ELT}

\author[1]{Rub\'en S\'anchez-Janssen}
\affil[1]{STFC UK Astronomy Technology Centre (UK)}
\author[2]{Francois Hammer}
\affil[2]{GEPI, Observatoire de Paris, PSL University, CNRS, France}
\author[3]{Simon Morris}
\affil[3]{Durham University, UK}
\author[4]{Jean-Gabriel Cuby}
\affil[4]{LAM, Université Aix-Marseille, CNRS, France}
\author[5,19]{Lex Kaper}
\affil[5]{University of Amsterdam, Netherlands}
\author[6]{Matthias Steinmetz}
\affil[6]{Leibniz-Institut fuer Astrophysik Potsdam, Germany}
\author[7]{Jose Afonso}
\affil[7]{IACE, Universidade de Lisboa, Portugal}
\author[8]{Beatriz Barbuy}
\affil[8]{IAG, Sao Paulo, Brazil}
\author[9]{Edwin Bergin} 
\affil[8]{University of Michigan, USA}
\author[10]{Alexis Finoguenov} 
\affil[10]{University of Helsinki, Finland}
\author[11]{Jesus Gallego} 
\affil[11]{Universidad Complutense de Madrid, Spain}
\author[12]{Susan Kassin} 
\affil[12]{Space Telescope Science Institute, USA}
\author[9]{Chris Miller}
\author[13]{Goran \"Ostlin} 
\affil[13]{Stockholm University, Sweden}
\author[14]{Laura Pentericci} 
\affil[14]{INAF, Observatorio Astronomico di Roma, Italy}
\author[15]{Daniel Schaerer} 
\affil[15]{University of Geneva, Switzerland}
\author[16]{Bodo Ziegler} 
\affil[16]{Vienna University, Austria}
\author[2]{Fanny Chemla} 
\author[17,21]{Gavin Dalton}
\affil[17]{STFC RAL Space, UK}
\author[2]{Fatima De Frondat} 
\author[1]{Chris Evans} 
\author[4]{David Le Mignant} 
\author[2]{Mathieu Puech} 
\author[2]{Myriam Rodrigues} 
\author[2]{Sylvestre Taburet} 
\author[4]{Lidia Tasca} 
\author[2]{Yanbin Yang} 
\author[2]{Sandrine Zanchetta} 
\author[3]{Timothy Butterley} 
\author[18]{Jean-Marc Conan} 
\affil[18]{ONERA, France}
\author[4]{Kjetil Dohlen}
\author[3]{Marc Dubbeldam} 
\author[4]{Kacem El Hadi} 
\author[19]{Annemieke Janssen}
\affil[19]{NOVA, Netherlands}
\author[4,18]{Thierry Fusco} 
\author[6]{Andreas Kelz} 
\author[20]{Marie Larrieu} 
\affil[20]{IRAP, Université de Toulouse, CNRS, France}
\author[21]{Ian Lewis}
\affil[21]{Oxford University, UK}
\author[18]{Caroline Lim}
\author[1]{Mike MacIntosh} 
\author[3]{Tim Morris} 
\author[19]{Ramon Navarro} 
\author[21]{Walter Seifert}
\affil[21]{LSW Heidelberg, Germany}

\begin{document} 
\maketitle

\begin{abstract}
MOSAIC is the planned multi-object spectrograph for the 39m Extremely Large Telescope (ELT). 
Conceived as a multi-purpose instrument, it offers both high multiplex and multi-IFU capabilities at a range of intermediate to high spectral resolving powers in the visible and the near-infrared. 
MOSAIC will enable unique spectroscopic surveys of the faintest sources, from  the oldest stars in the Galaxy and beyond to the first populations of galaxies that completed the reionisation of the Universe--while simultaneously opening up a wide discovery space.
In this contribution we present the status of the instrument ahead of Phase B, showcasing the key science cases as well as introducing the updated set of top level requirements and the adopted architecture.
The high readiness level will allow MOSAIC to soon enter the construction phase, with the goal to provide the ELT community with a world-class MOS capability as soon as possible after the telescope first light.
\end{abstract}


\section{INTRODUCTION}
\label{sec:intro}
The construction of a multi-object spectrograph (MOS) is planned as part of the ELT Instrumentation Roadmap. 
As detailed below, there is a vast range of science topics for such a workhorse instrument on the ELT, ranging from spectroscopy of faint stars in our Galaxy and the Local Group, to galaxy evolution from the present to the epoch of reionisation, as well as producing a cosmic inventory of baryonic and dark matter. 

The MOSAIC Consortium completed the instrument Phase A study in 2018\,\cite{Morris2018}. The baseline design\,\cite{Jagourel2018} combines the advantages of a highly-multiplexed visible and near-infrared spectrograph targeting numerous unresolved sources (High Multiplex Modes; HMM hereafter) with a mode featuring a more modest multiplex but that is able to spatially resolve extended sources (multi-IFU modes).
Following the review board recommendations the Consortium, in close collaboration with ESO and the community, set to evaluate the scientific capabilities of the baseline concept while also taking into account its complementarity with the other ELT instruments already under development.

As a result of this analysis it was deemed very important to strive to maximise the instrument efficiency by prioritising the HMM multiplex and providing as broad a simultaneous wavelength coverage as possible in all modes--all while also maintaining the high resolving power ($R > 18,000$) capabilities in both visible and infrared wavelengths.
Additionally, it was concluded that the NIR multi-IFU would benefit from wider IFU FOVs, whereas the visible multi-IFU mode was not considered of top priority in the 2030 landscape and has been dropped from the baseline architecture.
Finally, MOSAIC will aim at increasing its observing efficiency even further by enabling simultaneous visible and near-infrared observations whenever possible. 

In this contribution we present the new set of proposed top-level requirements (TLR), and detail the technical feasibility studies that are currently underway.
The manuscript is structured as follows.
In Section\,\ref{sec:sci} we provide a summary of the consolidated Science Cases that are driving the design of MOSAIC, while in Sections\,\ref{sec:tlr} and \ref{sec:concept} we describe the implications and challenges of the revised TLRs and the adopted technical solutions.
Section\,\ref{sec:final} presents the next steps of the project as it heads into Phase B.

\section{THE SCIENCE CASE FOR MOSAIC}
\label{sec:sci}  
The set of initial Science Cases for MOSAIC was distilled from the ELT-MOS White Paper \cite{Evans2015a}, and informed the TLRs for the Phase A study.
The cases have since been prioritised and developed further by the community and the Science Team, resulting in a number of consolidated Science Cases.  
Comprehensive simulations of the scientific performance of the instrument for several of these cases are summarised in Ref.\,\citenum{Puech2018}, and they are presented in full in Refs.\,\citenum{Disseau2014,Japelj2019,Puech2010,Puech2018} for SC1, SC2, SC3 and SC4, respectively.
The interested reader is referred to these references for more detail.

\subsection{SC1 -- First light galaxies and the reionization of the
Universe}
At present the inter-galactic medium (IGM) is fully ionised. Exactly when and how reionisation occurred is still unknown and a debated topic in modern astrophysics. The Planck 2018 results indicate that the IGM may have been already half-ionized by $z\sim9$, and the spectra of high-redshift quasars suggest that the reionisation process was completed around $z\sim6$.
The identification of the main ionising sources has been elusive until now due to their faintness. The overarching goal of this case is to derive a precise characterization of the ionisation state of the IGM during the first Gyr ($5 < z < 13$) after the Big Bang; to construct the timeline and topology of reionisation; and to observe the formation and growth of the first galaxies. 
MOSAIC will do so by carrying out very deep optical-to-near-IR spectroscopy to probe UV rest-frame absorption lines and nebular emission from high-z galaxies. With its unique combination of large telescope aperture, high sensitivity to extended emission and precise sky subtraction (see Section\,\ref{sec:tlr}) MOSAIC is the only instrument capable of enabling detailed studies of statistically-significant samples of $z\sim8$ Lyman-break galaxies and Ly$\alpha$ emitters.

\subsection{SC2 -- Inventory of matter}
One of the most ambitious goals of MOSAIC is to combine  visible and near-IR observations to conduct the first direct inventory of matter in the Universe at $z\sim3$, including:

\begin{itemize}
    \item To constrain the distribution of neutral gas in the IGM, by observing  the Ly$\alpha$ forest imprinted on the spectra of Lyman-break galaxies. This will allow us to reconstruct the three-dimensional density field of the IGM on Mpc scales to study its topology, its chemical properties, and to correlate galaxy positions with the density peaks.
    \item To probe all the gas phases in the circum-galactic medium (CGM). This can be done using projected galaxy pairs, where gas in the halo of the foreground source can leave detectable absorption line features in the spectrum of the background source. In addition, the combination of nebular emission line ratios and spatially-resolved morpho-kinematics from non-resonant lines enables to trace gas flows and determine the physical conditions of the CGM on $\sim$100 kpc scales.
    \item To characterise the dark matter profiles of disk galaxies, by measuring their ionised-gas rotation curves. These spatially-resolved observations will constrain the evolution of the stellar vs. halo mass relationship, of the star formation efficiency, and of disk and halo angular momentum.
\end{itemize}

\subsection{SC3 -- Mass assembly of galaxies through cosmic time}
Advancing the field of galaxy formation requires a comprehensive census of the mass assembly, star formation histories and stellar populations in galaxies.
MOSAIC will carry out chemo-dynamical studies or large samples of galaxies across a wide range of stellar masses and environments at $2<z<4$. 
Low-mass galaxies are of particular interest, because their shallow potential wells and bursty star formation histories make them excellent probes of the physics of stellar feedback.
MOSAIC will map the properties of the ionised haloes of starbursting dwarfs at $z\sim2$, as well as measure their internal kinematical structure. These systems are believed to be true analogs to high-z galaxies, and as such can be used to probe,  with unmatched detail and sensitivity, the physical conditions of these prominent contributors to the reionisation of the Universe.
Metallicity gradients and inhomogeneities are highly sensitive to the gas surface density, its kinematic structure (coherent rotation vs unordered motions), and the prevalence of inflows and outflows.

\subsection{SC4 -- Stellar populations beyond the Local Group}
MOSAIC on the ELT will advance the understanding of the star formation and chemical enrichment histories of galaxies out to Mpc distances through spectroscopic studies of their resolved stellar populations and star cluster systems.  Metallicity and kinematical studies of large samples of old stars in the outer haloes of external galaxies (e.g. the Sculptor Group) will provide an extraordinary tool to characterise their past evolutionary history. 
High SNR optical spectroscopy of extragalactic massive stars will improve and refine our understanding of the role of environment on stellar evolution, particularly at low metallicities.
Detailed chemical abundances of globular clusters in the Local Universe, and in particular around dwarf galaxies, will shed light on the chemical enrichment histories and the phenomenon of multiple stellar populations.
Other relevant cases include the kinematics and stellar populations of stellar streams surrounding nearby galaxies, and velocity dispersion measurements of dSphs out to distances of tens of Mpc.

\subsection{SC5 -- Galaxy archaeology}
The oldest and lowest metallicity stars that exist today carry the chemical imprint of the first massive objects that ended their lives as supernovae. 
MOSAIC will constrain the physics of early star formation by analysing the metallicity distribution function (MDF) in a variety of environments, from the Galactic bulge to the outer stellar halo, as well as in faint satellites. 
In the inner Galaxy, high resolution $H$-band observations will provide direct abundance estimates of iron-peak, $\alpha$-group, $s$-process and other light elements in globular clusters and field dwarfs--thus constraining the chemical enrichment history of the bulge. 
As for the faintest Galactic satellites ($L < 10^{5} L_{\odot}$), there are simply not enough giant branch stars in most of these systems to constrain their MDF or kinematics. Detailed chemodynamical studies of stars down to the main-sequence turn-off represent the only avenue to reveal the mystery of their true nature (star clusters or ultra-faint dwarf galaxies?).
Finally, a key goal of this SC is to study the Li abundance in metal-poor stars in other Local Group galaxies. Because the observed Li is likely primordial, the so-called Spite Plateau provides an estimate of the baryonic density of the Universe. 
High resolution spectroscopy ($R \approx 20,000$) is needed to confirm whether or not the plateau is universal.\\

Based on these SCs, we have recently embarked in the process of restructuring the MOSAIC Science Team into the correspoding Science Working Groups (SWGs). The main goal of the SWGs is to identify and define, together with the Project Scientists, the MOSAIC Design Reference Programmes (DRP). These are high-impact science programmes that span a range of topics and that are uniquely possible with MOSAIC. The DRPs will not only showcase the scientific potential of the instrument, but will also play a critical in identifying where the challenges lie—scientific, engineering, operational, software or otherwise.
It is expected that many of these DRPs will form the basis for future surveys with MOSAIC, either as part of the Guaranteed Time Observations, separate Large Programmes, and/or Public Surveys (see Ref.\,\citenum{Puech2018}).

\section{UPDATED TOP-LEVEL REQUIREMENTS}
\label{sec:tlr}  

\begin{table}[ht] 
\label{tab:tlr}
\caption{MOSAIC top-level requirements. The top (bottom) row in each cell corresponds to TLRs for the low (high) spectral resolution settings.} 
\begin{center}    
\begin{tabular}{
!{\vrule width 1.2pt}
l!{\vrule width 1.2pt}
c!{\vrule width 1.2pt}
c!{\vrule width 1.2pt}
c!{\vrule width 1.2pt}
}
\noalign{\hrule height 1.2pt}
\rule[-1ex]{0pt}{3.5ex} TLR & HMM-VIS & HMM-NIR & multi-IFU \\

\noalign{\hrule height 1.1pt}

\rule[-1ex]{0pt}{3.5ex} FOV & 40 & 40 &  40 \\
\rule[-1ex]{0pt}{0ex}    [arcmin$^2$]      &   &   &  \\
\noalign{\hrule height 1.1pt}

\rule[-1ex]{0pt}{3.5ex} Multiplex & 200 & 140 & 8 \\
\cline{2-2} \cline{3-2} \cline{4-2} 
\rule[-1ex]{0pt}{3.5ex}    [\#]      & 70  &  140 & 8 \\
\noalign{\hrule height 1.1pt}

\rule[-1ex]{0pt}{3.5ex} Resolving power & 4,000  & 4,000 & 4,000   \\
\cline{2-2} \cline{3-2} \cline{4-2} 
\rule[-1ex]{0pt}{3.5ex} [R]     & 18,000 & 9,000 \& 18,000 & 9,000 \& 18,000 \\
\noalign{\hrule height 1.1pt}

\rule[-1ex]{0pt}{3.5ex} Spectral coverage & 0.45-0.8  & 0.8-1.8 & 0.8-1.8   \\
\cline{2-2} \cline{3-2} \cline{4-2} 
\rule[-1ex]{0pt}{3.5ex}   [$\mu$m]   & 0.51-0.57 \& 0.61-0.67\,$^{a}$ & 0.77-0.89 \& 1.52-1.62\,$^{a}$ & 0.77-0.89 \& 1.52-1.62\,$^{a}$ \\
\noalign{\hrule height 1.1pt}

\rule[-1ex]{0pt}{3.5ex} Simultaneous bandwidth & 0.175  & 1.0 & 1.0   \\
\cline{2-2} \cline{3-2} \cline{4-2} 
\rule[-1ex]{0pt}{3.5ex}   [$\mu$m]   & 0.06 & 0.12 \& 0.10 & 0.12 \& 0.10 \\
\noalign{\hrule height 1.1pt}

\rule[-1ex]{0pt}{3.5ex} Subfield aperture & 0.7  & 0.5 & 2.5   \\
\rule[-1ex]{0pt}{0ex}   [arcsec]   & & & \\
\noalign{\hrule height 1.1pt}

\rule[-1ex]{0pt}{3.5ex} Spaxel size  & --  & -- & $<$ 200   \\
\rule[-1ex]{0pt}{3.5ex}   [mas]   & & & \\
\noalign{\hrule height 1.1pt}

\end{tabular}
\end{center}
\footnotesize{$^a$ Not yet fully optimised}\\
\end{table}

\begin{figure} [ht]
\begin{center}
\begin{tabular}{c}
\includegraphics[]{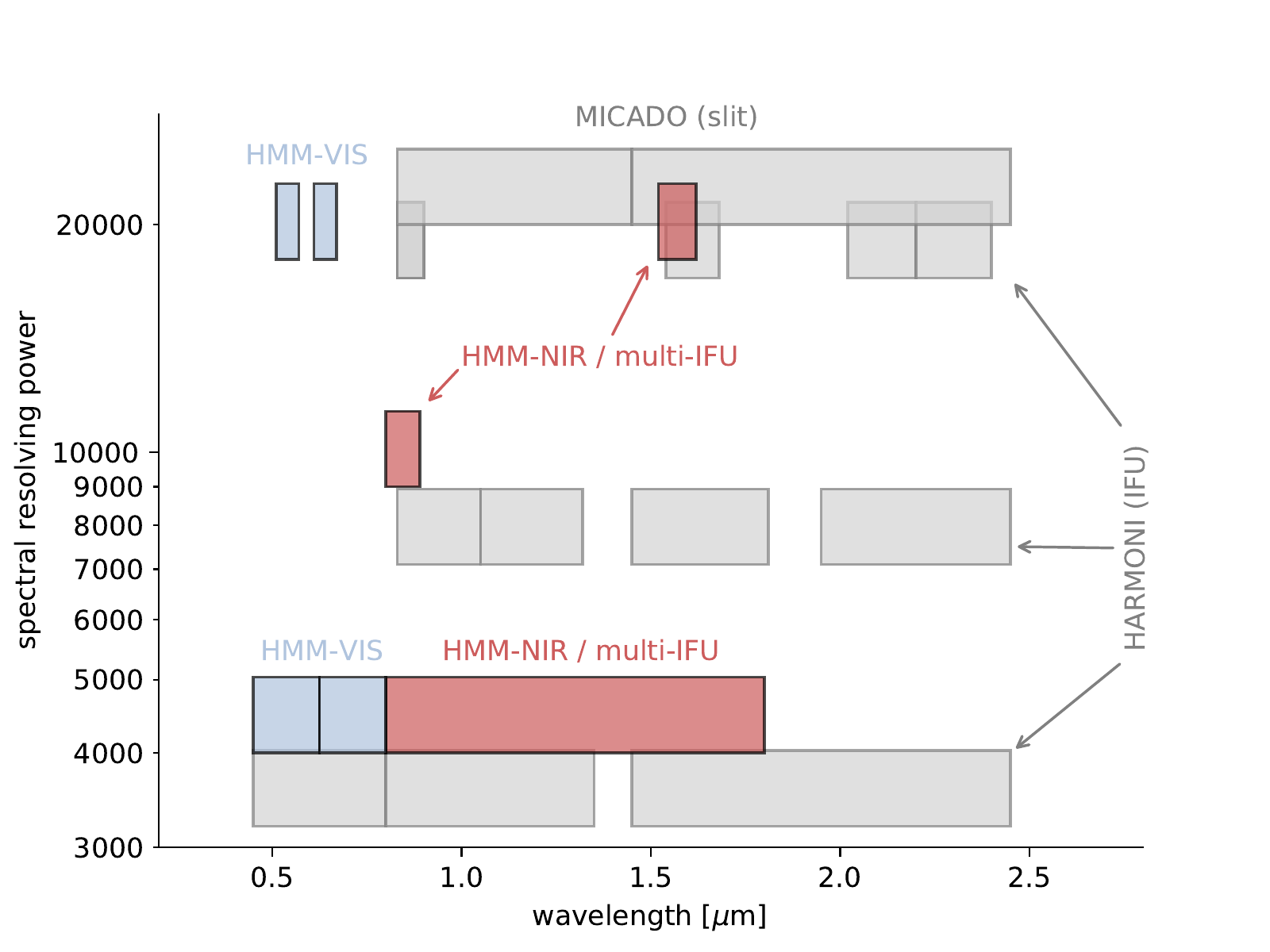}
\end{tabular}
\end{center}
\caption[example] 
{ \label{fig:wv} 
Spectral resolving power and wavelength coverage of the MOSAIC modes (in colour). For completeness, we also show the spectral settings of HARMONI and MICADO, the other ELT instruments offering intermediate resolving powers in the visible and near-infrared. MOSAIC will provide unique and highly complementary capabilities to the ELT instrumentation suite. See text for details.}
\end{figure}

Table\,\ref{tab:tlr} presents the set of proposed instrument TLRs for the next stages of the project, which flow down directly from the key Science Cases described in Section\,\ref{sec:sci}. 
Cells that contain two rows indicate the corresponding TLRs for the low and high resolution modes at the top and bottom, respectively.

The overarching principle behind these TLRs is to deliver a highly efficient HMM mode, which as a result should largely drive the  design of the instrument.
In addition to high multiplex and wide simultaneous bandwidth, coverage of visible wavelengths at a range of intermediate to high resolving powers is programmatically important for the ELT--as it is that the instrument retains a high throughput across a range of atmospheric conditions. 
In Fig.\ref{fig:wv} we  show the spectral settings of MOSAIC in the context of the other ELT instruments offering intermediate resolving powers in the same wavelength range, namely HARMONI\,\cite{Thatte2016} and MICADO\,\cite{Davies2018}.
It is clear that MOSAIC fills an important niche by providing a unique variety of spectroscopic modes. \cite{marconi2018}
\,\footnote{The ELT will also offer $R=100,000$ spectroscopy in the VIS+NIR and MIR through HIRES\,\cite{marconi2018} and METIS\,\cite{Brandl2018}, respectively. They are not represented in the figure for readability reasons.}

A number of the TLRs still require further optimisation, including the spectral bandwidths of the high resolution settings.
Additionally, the exact multiplex in HMM-NIR is still to be confirmed. 
This depends on whether or not the fibres can be deployed independently or have to be paired at the hardware level so as to carry out cross-beam switching\,\cite{rodrigues2012} to improve the sky subtraction accuracy.
In the latter case the effective multiplex will reduce to 70. 
Achieving the most precise sky subtraction is a major driver for MOSAIC, and the multi-IFU mode enables local measurements of the sky continuum that should substantially cancel out the expected spatio-temporal variations.
The enlarged multi-IFU FOV (2.5 arcsec, from 1.9 arcsec at Phase A) will surely be beneficial in this area.

\section{INSTRUMENT CONCEPT}
\label{sec:concept}  

\begin{figure} [ht]
\begin{center}
\begin{tabular}{c}
\includegraphics[height=13cm]{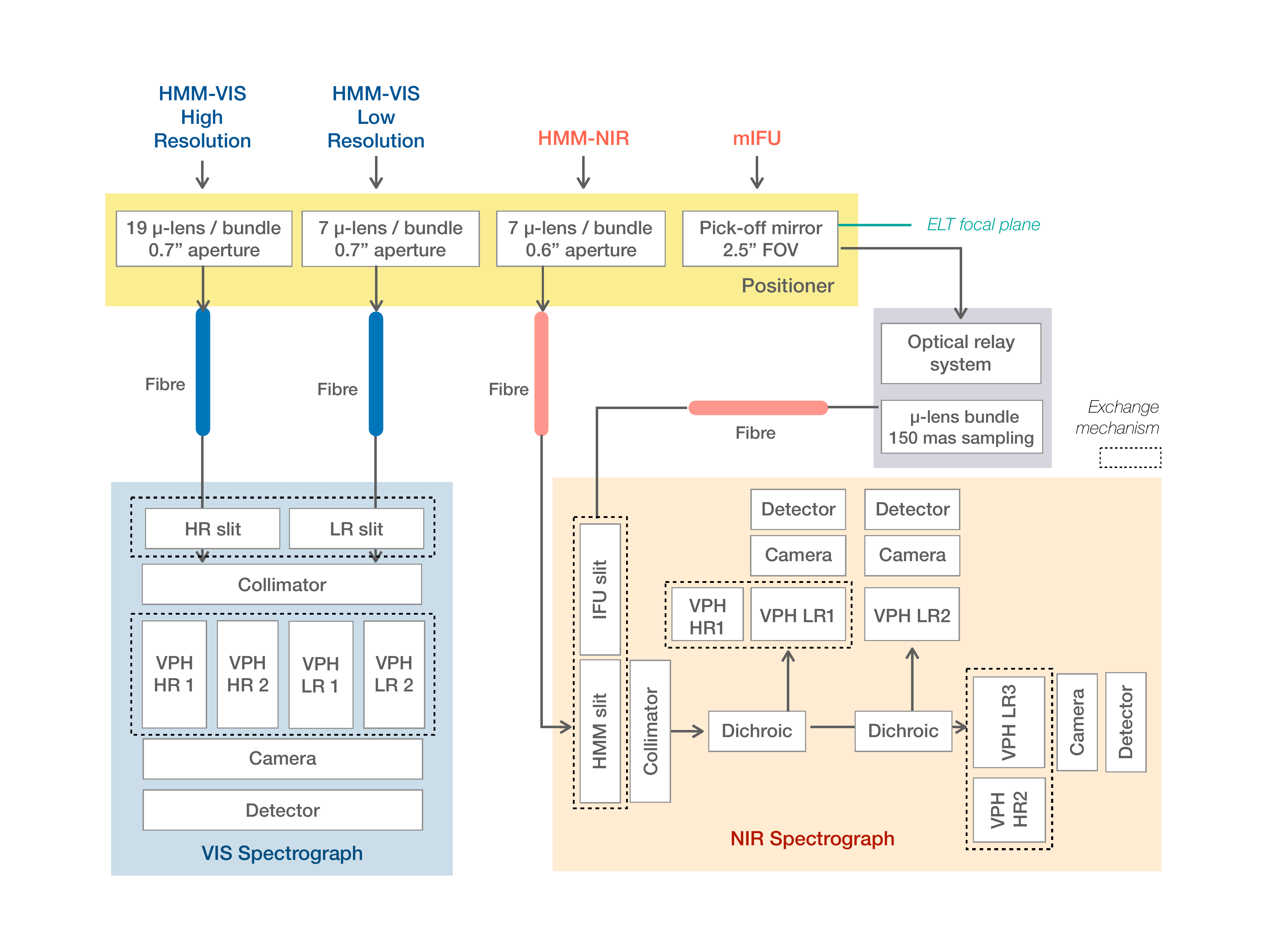}
\end{tabular}
\end{center}
\caption[example] 
{ \label{fig:concept} 
Schematic representation of the instrument architecture.}
\end{figure}

A detailed description of the instrument conceptual design was presented in Ref.\,\citenum{Jagourel2018}. 
Here we provide a brief summary, focusing on the elements that have since evolved or are relevant to accommodate the updated TLRs. 
A schematic representation of the instrument architecture is shown in Fig.\,\ref{fig:concept}.

MOSAIC was conceived as a multi-purpose spectrograph for the ELT, and as discussed offers three observing modes: HMM-VIS, HMM-NIR, and multi-IFU.
The multiplex requirement in the HMM modes dictates that the spectrographs shall be fibre-fed, and the design is further simplified by allowing the HMM-NIR and multi-IFU modes to share the same spectrograph. 
As per the updated TLRs, the three instrument modes operate between 0.45-1.8 $\mu$m, with a break at 0.8 $\mu$m between a system optimised for the visible and one optimised for the near-infrared. 
As shown in Fig.\ref{fig:wv}, the full spectral range is covered in three bands at $R=4,000$, with selected high resolution settings in the VIS ($R>18,000$ in two bands) and the NIR ($R > 9,000$ in the CaT region, and $R > 18,000$ in the $H$-band).

MOSAIC is the only ELT instrument that adopts smart focal plane technologies\,\cite{Cunningham2009} to make full use of the available AO-corrected FOV (7.5 arcmin).
Sub-fields can be allocated in a 10 arcmin diameter field, except in the regions vignetted by the four laser guide stars (LGS).
A substantial development is that MOSAIC is no longer planning to implement MOAO correction.
This is largely driven by the coarser spaxel size of the multi-IFU units, but also by a desire to reduce complexity and cost.
The AO system still provides Ground Layer Adaptive Optics (GLAO) correction across the full wavelength range of the instrument.
Preliminary simulations indicate that in the NIR and over excellent seeing conditions, GLAO can potentially reach MOAO performance levels, e.g., 25\% EE within 150 mas at 1.65 $\mu$m across the full FOV.
The performance at the blue end ($<0.6~\mu$m) is expected to be nearly equivalent to seeing-limited observations.
More comprehensive simulations incorporating tomographic reconstruction are underway to assess the impact of LGS asterism diameter, turbulence profile and, critically, ELT M4 conjugation and dome turbulence.

Due to the non-telecentricity of the telescope design, the focal plane is stepped and subdivided into hexagonal tiles.
In the HMM modes the subfields, which consist of a microlens array coupled to fibres, are positioned in the telescope focal plane by a  positioner mechanism (see Sect.\,\ref{sec:modifications}). The light is then transported by the fibre cables to the slit entrance of the spectrograph. In the multi-IFU mode light is selected at the telescope focal plane by a pick-off mirror located in the appropriate tile. The beam is then redirected to an optical relay system at the edge of the instrument field. The target is re-imaged after a path compensation device into an IFU (microlens array plus fibres). The same system of pick-off mirrors and relays is used for the NGS.

\subsection{Modifications to the baseline design}
\label{sec:modifications}

In order to accommodate the updated TLRs a number of modifications to the baseline design need to be implemented, and the corresponding trade-off studies are underway. 
These mostly affect the optical design of both spectrographs, as well as the fibre and positioner subsystems.

The focal plane tiles need not only to accommodate additional dedicated HR-VIS fibre bundles to reach the required spectral resolution, but the possibility to enable simultaneous HMM-VIS and HMM-NIR observations through independent movement of the  positioners will be explored.
A novel solution for the positioner subsystem based on hexapod technologies is currently under investigation.
While there are no hardware limitations to the implementation of simultaneous HMM-VIS and multi-IFU observations, the operational feasibility of this combined mode is yet to be analysed.
This is an aspect of high priority for MOSAIC, as it would highly increase the observing efficiency and scientific return of the instrument. 

To enable full simultaneous spectral coverage in the NIR the optical design of the spectrographs shall be similar to that of the MOONS instrument,\,\cite{Taylor2018a} with three spectral channels using f/0.95 Schmidt cameras. 
Conservation of $A\Omega$ then leads to fibre bundles with roughly $0.6$ arcsec apertures sampled by 7 fibres.
This can be achieved with a fibre core of 150 $\mu$m, which for the multi-IFU mode translates into 150 mas spaxels. 

The requirement for high multiplex and wide simultaneous wavelength coverage at low resolution in the visible can be met with 7 fibres bundled together in a 0.7 arcsec aperture.
The high resolution setting, on the other hand, requires 19 fibres sampling the same bundle aperture.
Feasible optical designs for the visible spectrograph include both fast dioptric and catadioptric cameras.

\section{TOWARDS THE CONSTRUCTION PHASE}
\label{sec:final}

MOSAIC on the ELT will play a fundamental role in the scientific landscape of the 2030s.
It will be the reference instrument for deep pencil-beam surveys of the faintest sources, addressing important science topics where statistical samples are essential to advance the field. These range from spectroscopic surveys of resolved stellar populations in the Galaxy and beyond, to studies of the first populations of galaxies that gave rise to the reionisation of the Universe.
MOSAIC will enable these cases by offering both high multiplex and multi-IFU capabilities at a range of intermediate to high spectral resolving powers in the visible and the near-infrared.
As a result, the instrument discovery space is extensive, which is a highly desirable characteristic in the light of  the long history of serendipitous astronomical discoveries. MOSAIC not only is highly complementary to the other planned ELT instruments, but it also provides very strong synergies with future facilities, e.g., by enabling detailed physical characterisation of faint JWST sources, or by providing follow-up visible and near-infrared spectroscopy for surveys at other wavelengths (e.g., SKA, ATHENA).

The MOSAIC Consortium is working closely with ESO with the aim to formalise the start of Phase B during the second half of 2021.  
As detailed in Section\,\ref{sec:modifications}, a number of technical feasibility studies are underway to sufficiently de-risk the proposed modifications to the instrument concept--but we note that the adopted architecture is largely based on proven designs and therefore already is of high technology readiness levels. 
This will allow MOSAIC to soon move into the next design stage, with the goal to provide the ELT community with a world-class MOS capability as soon as possible after the telescope first light.

\acknowledgments 
The MOSAIC Consortium is very grateful to the MOSAIC Science and Technical Teams for their essential contributions to the development of the instrument.
We also acknowledge the significant contribution of the MOSAIC Team at ESO and the community scientists to the refinement of the instrument TLRs.

\bibliography{library.bib} 
\bibliographystyle{spiebib} 

\end{document}